\documentclass[aps,12pt,preprint]{revtex4}

\usepackage{subeqnarray}

\begin{document}
\preprint{SNUTP02-030}
\title{\hspace{1cm}\\\hspace{1cm}\\\hspace{1cm}\\Stationary Perturbation
Theory with Spatially Well-separated Potentials}

\author{\hspace{1cm}\\Seok Kim}\email{calaf2@snu.ac.kr}
\author{Choonkyu Lee}\email{cklee@phya.snu.ac.kr}

\affiliation{\vspace{0.3cm}School of Physics and Center for
Theoretical Physics\\ Seoul National University, Seoul 151-747,
Korea \vspace{0.5cm}}

\begin{abstract}
We present a new perturbation theory for quantum mechanical energy
eigenstates when the potential equals the sum of two localized,
but not necessarily weak potentials $V_{1}(\vec{r})$ and
$V_{2}(\vec{r})$, with the distance $L$ between the respective
centers of the two taken to be quite large. It is assumed that
complete eigenfunctions of the local Hamiltonians (i.e., in the
presence of $V_{1}(\vec{r})$ or $V_{2}(\vec{r})$ only) are
available as inputs to our perturbation theory. If the two local
Hamiltonians have degenerate bound-state energy levels, a
systematic extension of the molecular orbital theory (or the
tight-binding approximation) follows from our formalism. Our
approach can be viewed as a systematic adaptation of the multiple
scattering theory to the problem of bound states.
\end{abstract}

\maketitle

\section{introduction}

In one-particle quantum mechanics, consider the energy eigenvalue
problem
\begin{equation}\label{sch-eq}
  \hat{H}|\psi\rangle=E|\psi\rangle,
\end{equation}
when the Hamiltonian of the system has two separate potential
contributions, viz.,
\begin{equation}\label{tot-ham}
  \hat{H}=\frac{1}{2m}\hat{p}^{2}+\hat{V}_{1}+\hat{V}_{2}.
\end{equation}
(For simplicity, we will present our discussion within a
one-dimensional context). Then one may also consider the related
eigenvalue problems
\begin{equation}\label{loc-sch-eq}
\begin{array}{c}
  \hat{H}_{1}|\psi_{1}\rangle=\varepsilon|\psi_{1}\rangle,\\
  \hat{H}_{2}|\psi_{2}\rangle=u|\psi_{2}\rangle,
\end{array}
\end{equation}
where
\begin{equation}\label{loc-ham}
  \hat{H_{1}}=\frac{1}{2m}\hat{p}^{2}+\hat{V}_{1}\ \ ,\ \ \
  \hat{H_{2}}=\frac{1}{2m}\hat{p}^{2}+\hat{V}_{2}\ .
\end{equation}
Generally speaking, there will be no simple connection between the
eigenvalue problem (\ref{sch-eq}) and those in (\ref{loc-sch-eq})
(other than an inequality-type relation for the ground state
energy). But, if $\hat{V}_{1}$ and $\hat{V}_{2}$ correspond to
some localized, but not necessarily weak, potentials with the
centers at $x=0$ and $x=L$, respectively and the separation
distance $L$ is relatively large, one might hope that the
solutions to the eigenvalue problems in (\ref{loc-sch-eq})
(involving `local Hamiltonians' $\hat{H_{1}}$ and $\hat{H_{2}}$)
be useful for generating good approximate solutions to the initial
problem (\ref{sch-eq}). Indeed, this view forms the basis of the
so-called molecular orbital theory or the tight-binding
approximation[1,2], in which one diagonalizes the full Hamiltonian
$\hat{H}$ within the truncated vector space given by a linear
combination of atomic orbitals (consisting of a few low-lying
eigenstates of the local Hamiltonians). In the context of
Born-Oppenheimer approximation where the local potential centers
are not really fixed, this kind of energy eigenvalue problem is of
particular importance since it can account for an effective
\textit{binding force} between the potential-producing objects.

The tight-binding method or its variants will be useful when given
local Hamiltonians allow some deeply-bound orbitals which are
separated from other local eigenstates by relatively large energy
gap. By its very nature, however, a reliable theoretical error
estimate for the scheme (especially when the parameters in the
given problem are not quite in the limiting range for the method)
is difficult to make. Also, if one of the local potentials, say,
$\hat{V}_{2}$, happens to be strictly repulsive (and so no atomic
orbital associated with $\hat{H}_{2}$) while $\hat{H}_{1}$ allows
some bound states, this method is unable to give any useful
information on the effect of the potential $\hat{V}_{2}$ on the
low-lying eigenstates of the full Hamiltonian (\ref{tot-ham}).
There is a related question within the usual tight-binding
approximation, that is, on the role of the \textit{continuum
states} in the scheme. Clearly, it is desirable to have a
systematic approximation scheme which goes beyond the simplest
tight-binding approach. [Recently, Barton et al.[3] discussed the
effect of a distant impenetrable wall on quantum mechanical energy
levels; but their approach is tuned to the change of the boundary
condition, and therefore does not apply to more generic case
involving two well-separated potentials.]

In this paper we develop a new stationary perturbation theory
which can be used to study the eigenvalue problem with a
two-centered Hamiltonian. (For a Hamiltonian with more than two
centers a simple extension of our method should be useful.) While
there exists a systematic theory dealing with scattering by a
multi-centered potential (see Ref.[4] for instance), we are not
aware of such development which can be used to study the
corresponding bound-state problem in a well-controlled manner. In
our approach to the eigenvalue problem (\ref{sch-eq}), it will be
assumed that the eigenvalue problems with the local Hamiltonians
can be solved explicitly, and so we have at our disposal a
complete orthonormal set $\{|n\rangle\}$ based on eigenstates of
$\hat{H}_{1}$ and another complete orthonormal set
$\{|\bar{n}\rangle\}$ based on eigenstates of $\hat{H}_{2}$. [The
knowledge of the Green's operators associated with the local
Hamiltonians may be assumed instead.] We wish to exploit this
\textit{over-complete} set of basis, which include continuum
states, in constructing the bound states of the total Hamiltonian
$\hat{H}$. The result is a perturbation series in which the
expansion parameter is a quantity approaching zero as the
separation between the local potentials becomes large. [In fact,
for strongly localized local potentials, we have an expansion
parameter of order $e^{-\alpha L}$ ($\alpha$: constant)]. It can
be viewed as an expansion in the \textit{wave-function stretching
factor}, that comes with for every $\hat{V}_{2}$ ($\hat{V}_{1}$)
acting on a specific bound state of  $\hat{H}_{1}$
($\hat{H}_{2}$). This small factor is a direct measure on how much
influence one local potential feels from the bound states
associated with the other local potential. We also remark that the
general philosophy of our formalism is similar to that of the
multiple scattering theory\cite{gold}, but the very nature of the
bound-state eigenvalue problems necessitates somewhat different
developments.

It should be noted that the standard time-independent perturbation
theory is generally unreliable for our problem. To see that, it
suffices to consider the simple situation where one has the bound
state energy levels of $\hat{H}_{1}$ influenced by a strictly
positive, well-localized potential $V_{2}(x)$ at a large distance
$L$ (from the center of the potential $V_{1}(x)$). The strength of
$\hat{V}_{2}$ may not be small, however. If this case can be
studied by the usual perturbation theory (that is, by treating
$\hat{V}_{2}$ as a perturbation to the unperturbed Hamiltonian
$\hat{H}_{1}$), the state $|k\rangle$ satisfying
$\hat{H}_{1}|k\rangle=\varepsilon_{k}|k\rangle$, i.e., that with
the unperturbed energy $E^{(0)}=\varepsilon_{k}\ (<0)$, would
acquire the first- and second-order energy shifts
\begin{subeqnarray}
&&\hspace{-1cm}E^{(1)} = \langle k|\hat{V}_{2}|k\rangle\slabel{naive-1st},\\
&&\hspace{-1cm}E^{(2)} =
        \sum_{n(\neq k)}
        \frac{\langle k|\hat{V}_{2}|n\rangle \langle
        n|\hat{V}_{2}|k\rangle}{\varepsilon_{k}-\varepsilon_{n}}\
        ,\slabel{naive-2nd}
\end{subeqnarray}
assuming for simplicity no degeneracy for the unperturbed states.
According to (\ref{naive-1st}), $E^{(1)}$ would be of order
$e^{-\frac{2}{\hbar}\!\sqrt{2m|\varepsilon_{k}|}\ L}$. This is
nothing but the product of two wave-function stretching factors,
as appropriate to the matrix element of $\hat{V}_{2}$ in a
specific bound state of $\hat{H}_{1}$. The fact is that, according
to (\ref{naive-2nd}), $E^{(2)}$ would also be
${\mathcal{O}}(e^{-\frac{2}{\hbar}\!\sqrt{2m|\varepsilon_{k}|}\
L})$ due to the continuum contribution in the intermediate-state
sum. This implies that, depending on the strength of
$\hat{V}_{2}$, the second order shift $E^{(2)}$ might be as big as
the first-order shift. In an analogous manner, it is not difficult
to see that the contributions from the continuum states make the
r-th order shift $E^{(r)}$ assume the same order of magnitude as
$E^{(1)}$. Hence this is not a valid expansion, and we have to
devise a more elaborate scheme to solve our problem.

This paper is organized as follows. In Sec.2, we will concentrate
on setting up a reliable perturbation theory with two spatially
well-separated potentials in the nondegenerate case. In this
discussion we will suppose (mainly to have mathematics under
control) that the potentials $V_{1}$ and $V_{2}$ are sufficiently
well localized; but, we expect that most of our formulas, with
suitable adjustments if necessary, remain useful even if these
potentials are localized only by some (not too small) powers in
the distance from the respective potential centers. Our method is
exhibited explicitly for local Hamiltonians involving
$\delta-$function potentials. Section 3 is devoted to the
extension of this method to the case where the local Hamiltonians
$\hat{H}_{1}$ and $\hat{H}_{2}$ have (almost-)degenerate energy
levels. Here one sees explicitly that, for a reliable perturbation
series, a separate treatment in the subspace of degenerate local
bound states becomes necessary. The resulting theory is a
generalization of the molecular orbital theory that allows one to
systematically study higher order corrections, and as such it
should have some practical value as well. Section 4 contains
concluding remarks. In the Appendix we present our argument behind
the order estimates for various contributions appearing in our
perturbation theory (together with some analysis for the example
problem).

\section{nondegenerate perturbation theory}

Our goal is to obtain approximate eigenstates of the Hamiltonian
$\hat{H}=\frac{1}{2m}\hat{p}^{2}+\hat{V}_{1}+\hat{V}_{2}$, when
complete solutions to the eigenvalue problems with the local
Hamiltonians
$\hat{H_{1}}(\equiv\frac{1}{2m}\hat{p}^{2}+\hat{V}_{1})$ and
$\hat{H_{2}}(\equiv\frac{1}{2m}\hat{p}^{2}+\hat{V}_{2})$ are
known. In the Hilbert space ${\mathcal{V}}$ of the system we have
with us two complete orthonormal sets --- the set $\{|n\rangle\}$
based on (discrete and continuous) eigenstates of  $\hat{H}_{1}$
and the set $\{|\bar{n}\rangle\}$ based on eigenstates  of
$\hat{H}_{2}$. Let $|k\rangle$ be a given specific nondegenerate
bound state of $\hat{H}_{1}$, with eigenvalue $\varepsilon_{k}$.
We further assume in this section that no eigenstate of
$\hat{H}_{2}$ has the eigenvalue equal or very close to
$\varepsilon_{k}$. Then, if the distance $L$ between the centers
of two local potentials $V_{1}(x)$ and $V_{2}(x)$ is large enough,
we expect that the full Hamiltonian $\hat{H}$ admit an energy
eigenstate $|\phi_{k}\rangle$ which should coincide with
$|k\rangle$ in the limit $L\rightarrow\infty$ (i.e., as $V_{2}(x)$
is sent away to the very remote). This should be the case
irrespectively of the relative magnitude of the two local
potentials. Thus, for large $L$, we may write the solution to the
eigenvalue equation
\begin{equation}\label{original-sch}
  \hat{H}|\phi_{k}\rangle = E_{k}|\phi_{k}\rangle\ \ ,\ \ \
  (\hat{H}=\hat{H}_{1}+\hat{V}_{2})
\end{equation}
as
\begin{equation}
  E_{k} = \varepsilon_{k}+\delta E_{k}\ \ ,\ \ \
  |\phi_{k}\rangle = |k\rangle+|\delta\phi_{k}\rangle.
\end{equation}
Here the small corrections $\delta E_{k}$ ,
$|\delta\phi_{k}\rangle$ should satisfy the equation
\begin{equation}\label{sch-decomp}
  (\varepsilon_{k}-\hat{H}+\delta E_{k})|\delta\phi_{k}\rangle
  = (\hat{V}_{2}-\delta E_{k})|k\rangle,
\end{equation}
which is still exact.

Let us now study the implication of (\ref{sch-decomp}) in detail.
First of all, as in the ordinary stationary perturbation theory,
(\ref{sch-decomp}) does not determine $|\delta\phi_{k}\rangle$
uniquely[5]: if $|\delta\phi_{k}\rangle$ is a solution of
(\ref{sch-decomp}), so is
$|\delta\phi_{k}\rangle'=\frac{1}{1+\beta}\{|\delta\phi_{k}\rangle-\beta|
k\rangle\}$ for arbitrary constant $\beta$. As a result, $\langle
k|\delta\phi_{k}\rangle$ may be chosen as one wishes and the
particularly convenient, at least in the ordinary perturbation
theory, is the choice
\begin{equation}\label{compliment}
  \langle k|\delta\phi_{k}\rangle=0,
\end{equation}
i.e., define $|\delta\phi_{k}\rangle$ in the subspace
${\mathcal{V}_{k}}$, the orthogonal complement of $|k\rangle$ in
${\mathcal{V}}$. With this choice and multiplying both sides of
(\ref{sch-decomp}) by $\langle k|$ on the left, one obtains
\begin{equation}\label{shift-formula}
  \delta E_{k} = \langle k|\hat{V}_{2}|k\rangle + \langle
                 k|\hat{V}_{2}|\delta\phi_{k}\rangle\ \
  \left( = \langle k|\hat{V}_{2}|\phi_{k}\rangle\right).
\end{equation}
At the same time, one may replace (\ref{sch-decomp}) by
\begin{equation}\label{projected-sch}
  \hat{Q_{k}}(\varepsilon_{k}-\hat{H}+\delta E_{k})|\delta\phi_{k}\rangle
  = \hat{Q}_{k}\hat{V}_{2}|k\rangle,
\end{equation}
where $\hat{Q}_{k}\equiv 1-|k\rangle\langle k|$. Note that,
without the knowledge on $|\delta\phi_{k}\rangle$, the formula
(\ref{shift-formula}) is not informative by itself. To have
$|\delta\phi_{k}\rangle$ determined, one might write (as in the
conventional perturbation theory)
$|\delta\phi_{k}\rangle=\sum_{n(\neq k)}|n\rangle\langle
n|\delta\phi_{k}\rangle$ and determine $\langle
n|\delta\phi_{k}\rangle$ with the help of the equations resulting
from multiplying (\ref{projected-sch}) by $\langle n|$ on the
left. But, as was explained in the introduction, this usual
procedure does not lead to a useful perturbation series. (See also
discussions further below.)

At this point, recall that, in association with the second local
Hamiltonian $\hat{H}_{2}$, we have another complete set
$\{|\bar{n}\rangle\}$ where
$\hat{H}_{2}|\bar{n}\rangle=u_{\bar{n}}|\bar{n}\rangle$. We shall
utilize them with (\ref{sch-decomp}) in a suitable manner. (See
(\ref{free-prop}) below.) Here it is convenient to recast
(\ref{sch-decomp}) and (\ref{projected-sch}) as
\begin{equation}\label{sch-decomp-oper}
  \hat{O}|\delta\phi_{k}\rangle = \hat{V}_{2}|k\rangle-\delta
  E_{k}|k\rangle,
\end{equation}
\begin{equation}\label{proj-sch-decomp-oper}
  \hat{Q}_{k}\hat{O}|\delta\phi_{k}\rangle = \hat{Q}_{k}\hat{V}_{2}|k\rangle,
\end{equation}
introducing the operator
\begin{equation}\label{exact-prop}
  \hat{O} \equiv \varepsilon_{k}-\hat{H}+\delta E_{k}
  =\varepsilon_{k}-\hat{H}_{2}-(\hat{V}_{1}-\delta E_{k}).
\end{equation}
But we are not going to use the condition (\ref{compliment}) ---
it is not convenient for our development. [Note that
(\ref{projected-sch}) holds good without assuming this condition].
On $|\delta\phi_{k}\rangle$ we only demand that it should be
small, i.e., suppressed by at least one wave-function stretching
factor (accompanying, say, a term like $\hat{V}_{2}|k\rangle$). If
we multiply (\ref{sch-decomp-oper}) by $\langle k |$ on the left
\textit{without imposing} (\ref{compliment}), we obtain
\begin{equation}\label{gen-shift-formula}
  \delta E_{k} = \frac{\langle k|\hat{V}_{2}|k\rangle + \langle
                 k|\hat{V}_{2}|\delta\phi_{k}\rangle}
                 {1+\langle k|\delta\phi_{k}\rangle}\ \ .
\end{equation}
From this formula, we may conclude that $\delta E_{k}$ contains at
least two wave-function stretching factors. Now note that, when
$\hat{G}_{2}$ denotes the Green's operator associated with the
second Hamiltonian $\hat{H}_{2}$
\begin{equation}
  \hat{G}_{2}\equiv\frac{1}{\varepsilon_{k}-\hat{H}_{2}}
             =\sum_{\bar{n}}\frac{|\bar{n}\rangle\langle\bar{n}|}
                              {\varepsilon_{k}-u_{\bar{n}}}\label{free-prop}\ ,
\end{equation}
the operator $\hat{O}$ satisfies the relation
\begin{equation}
  1=\hat{O}\hat{G}_{2}+
            (\hat{V}_{1}-\delta E_{k})\hat{G}_{2},\label{it-rel}\ .
\end{equation}
Hence the right hand side of (\ref{proj-sch-decomp-oper}) may be
written as $\hat{Q}_{k}\{\hat{O}\hat{G}_{2}+(\hat{V}_{1}-\delta
E_{k})\hat{G}_{2}\}\hat{V}_{2}|k\rangle$, and then, by
rearranging, we obtain
\begin{equation}\label{1st-int-eq}
  \hat{Q}_{k}\hat{O}\left\{|\delta\phi_{k}\rangle
  -\hat{G}_{2}\hat{V}_{2}|k\rangle\right\}=
  \hat{Q}_{k}\hat{V}_{1}\hat{G}_{2}\hat{V}_{2}|k\rangle-
  \delta E_{k}\hat{Q}_{k}\hat{G}_{2}\hat{V}_{2}|k\rangle.
\end{equation}
This equation is the crucial one for our perturbation scheme.

We wish to solve (\ref{1st-int-eq}) order by order, with the order
in our case determined by the number of the wave-function
stretching factors involved. In its left hand side we have the
operator $\hat{O}$ acting on a vector yet to be found,
$|${\scriptsize${\mathcal{W}}$}$\rangle\equiv|\delta\phi_{k}\rangle-
\hat{G}_{2}\hat{V}_{2}|k\rangle$. Here it is important to note
that, if  $|${\scriptsize${\mathcal{W}}$}$\rangle$ does not
contain a component proportional to $|k\rangle$,
$|${\scriptsize${\mathcal{W}}$}$\rangle$ and
$\hat{O}|${\scriptsize${\mathcal{W}}$}$\rangle$ would be of the
same order due to the assumed nondegenerate nature of $\hat{H}$.
As for the component proportional to $|k\rangle$ from
$|${\scriptsize${\mathcal{W}}$}$\rangle$, on the other hand, the
situation is not the same: if $\hat{O}$ acts on that piece, the
resulting vector will have the order increased by at least one
wave-function stretching factor. This follows from
\begin{equation}
  \hat{O}|k\rangle=(\varepsilon_{k}-\hat{H}_{1}+\delta E_{k}-\hat{V}_{2})|k\rangle
  =(\delta E_{k}-\hat{V}_{2})|k\rangle.
\end{equation}
Based on this observation, the following conclusion should be
immediate: in (\ref{1st-int-eq}), the vector
$|${\scriptsize${\mathcal{W}}$}$\rangle$
($=|\delta\phi_{k}\rangle- \hat{G}_{2}\hat{V}_{2}|k\rangle$)
appearing in its left hand side is necessarily of the same order
as the expressions in its right hand side, under the proviso that
this restriction on the order does not apply to the term
proportional to $|k\rangle$. We here make another important
observation: the expressions we have in the right hand side of
(\ref{1st-int-eq}) are in fact of higher order than that of
$\hat{G}_{2}\hat{V}_{2}|k\rangle$. For its justification, see the
Appendix. Hence, setting
$|${\scriptsize${\mathcal{W}}$}$\rangle\approx 0$, i.e.,
$|\delta\phi_{k}\rangle- \hat{G}_{2}\hat{V}_{2}|k\rangle\approx 0$
solves (\ref{1st-int-eq}) to the leading order. We may thus write
\begin{equation}\label{1st-state}
  |\delta\phi_{k}\rangle^{(1)} = \hat{G}_{2}\hat{V}_{2}|k\rangle
\end{equation}
and, using this with (\ref{gen-shift-formula}), the following
formula for the energy shift results:
\begin{equation}\label{1st-shift}
  \delta E_{k}^{(1)} = \langle k|\hat{V}_{2}|k\rangle
                     + \langle k|\hat{V}_{2}\hat{G}_{2}\hat{V}_{2}|k\rangle\ .
\end{equation}
[Note that $\langle k|\delta\phi_{k}\rangle$ in the denominator of
(\ref{gen-shift-formula}) can be ignored in the leading-order
approximation.]

With the expression (\ref{free-prop}) for $\hat{G}_{2}$ inserted
into (\ref{1st-state}) and (\ref{1st-shift}), our formulas giving
the leading-order correction to the energy eigenstate may be
recast as the ones involving the sum over the basis set
$\{|\bar{n}\rangle\}$:
\begin{eqnarray}
  &&|\delta\phi_{k}\rangle^{(1)}=\sum_{\bar{n}}|\bar{n}\rangle\frac{
  \langle\bar{n}|\hat{V}_{2}|k\rangle}{\varepsilon_{k}-u_{\bar{n}}}\ ,
  \label{low-state-exp}\\
  &&\delta E_{k}^{(1)} = \langle k|\hat{V}_{2}|k\rangle
  + \sum_{\bar{n}}\frac{\langle k|\hat{V}_{2}|\bar{n}\rangle
  \langle\bar{n}|\hat{V}_{2}|k\rangle}{\varepsilon_{k}-u_{\bar{n}}}\
  .\label{low-shift-exp}
\end{eqnarray}
Note that, in our procedure, no explicit condition (like that in
(\ref{compliment})) has been used to dispense with the ambiguity
concerning the $|k\rangle$-component of $|\delta\phi_{k}\rangle$.
Instead, we have decided to choose the simplest available
expression for $|\delta\phi_{k}\rangle$, as suggested by the
order-by-order analysis of the relevant equation for our
perturbative development. In view of (\ref{1st-state}), one may
well say that our choice in fact corresponds to
\begin{equation}\label{compliment-2}
  \langle k|\delta\phi_{k}\rangle =
  \langle k|\hat{G}_{2}\hat{V}_{2}|k\rangle\ +\
  (higher\ order).
\end{equation}
The energy eigenstate we obtain is not properly normalized in
general.

Suppose one attacked the above problem with the help of the
conventional perturbation theory, regarding $\hat{V}_{2}$ as a
would-be perturbation. Then, instead of (\ref{1st-int-eq}), one
would work with the equation
\begin{equation}\label{1st-it-wrong}
  \hat{Q}_{k}\hat{O}(|\delta\phi_{k}\rangle-\hat{G}_{1}'\hat{V}_{2}|k\rangle) =
  \hat{Q}_{k}\hat{V}_{2}\hat{G}_{1}'\hat{V}_{2}|k\rangle -
  \delta E_{k}\hat{Q}_{k}\hat{G}_{1}'\hat{V}_{2}|k\rangle\ .
\end{equation}
where $\hat{G}_{1}'=\sum_{n(\neq k)}\frac{|n\rangle\langle
n|}{\varepsilon_{k}-\varepsilon_{n}}$ is the Green's operator
associated with the local Hamiltonian $\hat{H}_{1}$ (but defined
in the orthogonal complement ${\mathcal{V}_{k}}$). To obtain
(\ref{1st-it-wrong}), one may utilize the equation
(\ref{it-rel-1}) given below with (\ref{proj-sch-decomp-oper}).
The lowest-order approximation in the conventional perturbation
theory is tantamount to identifying $|\delta\phi_{k}\rangle$ with
the term $\hat{G}_{1}'\hat{V}_{2}|k\rangle$. But, in our case,
this is not a good approximation (unless the strength of
$\hat{V}_{2}$ itself is very weak), since the first term in the
right hand side of (\ref{1st-it-wrong}) can generate a comparable
contribution. (See the related discussion in the introduction).
Note that we had a different situation with (\ref{1st-int-eq}) ---
the expressions in its right hand side were of higher order (i.e.,
involved more wave-function stretching factors)!

To be convinced of the validity of our leading-order
approximations in (\ref{low-state-exp}) and (\ref{low-shift-exp}),
let us consider a simple example consisting of a pair of
$\delta$-function potentials, i.e.,
\begin{equation}\label{delta-pot}
V_{1}(x)=-\gamma_{1}\delta(x)\ ,\ V_{2}(x)=-\gamma_{2}\delta(x-L)
\end{equation}
with $\gamma_{1}>\gamma_{2}>0$. Then we know that each local
Hamiltonian admits one bound state. If $|1\rangle$ ($|2\rangle$)
denotes the bound state of $\hat{H}_{1}\equiv
\frac{1}{2m}\hat{p}^{2}+\hat{V}_{1}$ (of $\hat{H}_{2}\equiv
\frac{1}{2m}\hat{p}^{2}+\hat{V}_{2}$), the corresponding
(normalized) wave-function and energy eigenvalue read
\begin{eqnarray}
  &&\phi_{0}(x)\equiv\langle x|1\rangle=\sqrt{\frac{m\gamma_{1}}{\hbar^{2}}}\
  e^{-\frac{m\gamma_{1}}{\hbar^{2}}|x|}\ ,\
  ({\rm with}\ \varepsilon_{1}=-\frac{m\gamma_{1}^{2}}{2\hbar^{2}})\label{delta-1st-bound}\\
  &&\xi_{0}(x)\equiv\langle x|2\rangle=\sqrt{\frac{m\gamma_{2}}{\hbar^{2}}}\
  e^{-\frac{m\gamma_{2}}{\hbar^{2}}|x-L|}\ ,\
  ({\rm with}\ u_{2}=-\frac{m\gamma_{2}^{2}}{2\hbar^{2}})
\end{eqnarray}
respectively. Now, if the distance between the two local
potentials, $L$, is large (and the value of $u_{2}$ differs from
that of $\varepsilon_{1}$ significantly), we expect that the full
Hamiltonian
$\hat{H}=\frac{1}{2m}\hat{p}^{2}+\hat{V}_{1}+\hat{V}_{2}$ allow
two bound states $|\phi_{1}\rangle$ and $|\phi_{2}\rangle$, which
are approximately equal to $|1\rangle$ and $|2\rangle$,
respectively. For this example one can of course find the exact
bound state energies by a direct analysis of the corresponding
Schr\"{o}dinger equation. Explicitly, for the state
$|\phi_{1}\rangle$, its energy $E=-\frac{m\eta^{2}}{2\hbar^{2}}$
is determined by the equation
\begin{equation}\label{delta-energy-eq}
  \eta^{2}-(\gamma_{1}+\gamma_{2})\eta+\gamma_{1}\gamma_{2}
  (1-e^{-2\frac{mL}{\hbar^{2}}\eta})=0,
\end{equation}
and therefore, for large $L$, one has
\begin{equation}\label{delta-non-exact}
  E=-\frac{m\gamma_{1}^{2}}{2\hbar^{2}}\left\{1+
  \frac{2\gamma_{2}}{\gamma_{1}-\gamma_{2}}
  e^{-2\frac{m\gamma_{1}}{\hbar^{2}}L}+{\mathcal{O}}
  (e^{-4\frac{m\gamma_{1}}{\hbar^{2}}L})\right\}.
\end{equation}

The above result can also be obtained by using our formula
(\ref{low-shift-exp}). For such check, we need a complete basis
$\{|\bar{n}\rangle\}$ consisting of the energy eigenstates of
$\hat{H}_{2}$: in position space, the desired complete set
contains, aside from the bound state $\xi_{0}(x)$, two distinct
classes of continuum states (both corresponding to energy
$u_{q}=\frac{\hbar^{2}q^{2}}{2m}$)
\begin{equation}\label{delta-eigen}
 \begin{array}{l}\displaystyle
  \langle x|\bar{q}^{(1)}\rangle=\frac{1}{\sqrt{\pi}}\cos[q|x-L|+\tan^{-1}
  (\frac{m\gamma_{2}}{q\hbar^{2}})],\\
  \displaystyle\langle x|\bar{q}^{(2)}\rangle=\frac{1}{\sqrt{\pi}}\sin[q(x-L)].
 \end{array}
\end{equation}
Then, by straightforward calculations using these eigenfunctions,
we find
\begin{subeqnarray}
  ({\rm i})&&\!\!\langle1|\hat{V}_{2}|1\rangle=-2\frac{\gamma_{2}}{\gamma_{1}}
  \left(\frac{m\gamma_{1}^{2}}{2\hbar^{2}}\right)
  e^{-2\frac{m\gamma_{1}}{\hbar^{2}}L},\\
  ({\rm ii})&&\!\!\frac{\langle1|\hat{V}_{2}|2\rangle\langle2|\hat{V}_{2}|1\rangle}
  {\varepsilon_{1}-u_{2}}=-\frac{4(\frac{\gamma_{2}}{\gamma_{1}})}
  {(\frac{\gamma_{1}}{\gamma_{2}})^{2}\!-\!1}
  \left(\frac{m\gamma_{1}^{2}}{2\hbar^{2}}\right)
  e^{-2\frac{m\gamma_{1}}{\hbar^{2}}L},\\
  ({\rm iii})&&\!\!\int_{0}^{\infty}\!\!dq\frac{\langle1|\hat{V}_{2}|\bar{q}^{(1)}\rangle
  \langle\bar{q}^{(1)}|\hat{V}_{2}|1\rangle\!+\!\langle1|\hat{V}_{2}|\bar{q}^{(2)}\rangle
  \langle\bar{q}^{(2)}|\hat{V}_{2}|1\rangle}
  {\varepsilon_{1}-u_{q}}\!=\!-\frac{2(\frac{\gamma_{2}}{\gamma_{1}})}
  {(\frac{\gamma_{1}}{\gamma_{2}})\!+\!1}
  \left(\frac{m\gamma_{1}^{2}}{2\hbar^{2}}\right)\!
  e^{-2\frac{m\gamma_{1}}{\hbar^{2}}L}\!.
\end{subeqnarray}
By summing these contributions, we thus obtain the result $\delta
E^{(1)}=-\frac{2\gamma_{2}}{\gamma_{1}-\gamma_{2}}(\frac{m\gamma_{1}^{2}}{2\hbar^{2}})
e^{-2\frac{m\gamma_{1}}{\hbar^{2}}L}$, which is in agreement with
(\ref{delta-non-exact}). One may also calculate the first order
eigenfunction correction with the help of our formula
(\ref{low-state-exp}). After some straightforward calculations, we
then find
\begin{equation}
  \phi_{0}(x)+\langle x|\delta\phi_{1}\rangle^{(1)}=
  \sqrt{\frac{m\gamma_{1}}{\hbar^{2}}}\left\{
  e^{-\frac{m\gamma_{1}}{\hbar^{2}}|x|}+\frac{\gamma_{2}}{\gamma_{1}-\gamma_{2}}
  e^{-\frac{m\gamma_{1}}{\hbar^{2}}L}e^{-\frac{m\gamma_{1}}{\hbar^{2}}|x-L|}\right\} .
\end{equation}
This is the correct result, for the exact eigenfunction in the
limit $e^{-\frac{m\eta}{\hbar^{2}}L}\rightarrow0$ (with
$\eta=\sqrt{-\frac{2\hbar^{2}E}{m}}=\gamma_{1}+O(e^{-2\frac{m\gamma_{1}}{\hbar^{2}}L})$
from (\ref{delta-energy-eq})) can be approximated by
\begin{equation}
  \sqrt{\frac{m\gamma_{1}}{\hbar^{2}}}\left\{
  e^{-\frac{m\eta}{\hbar^{2}}|x|}+\frac{\gamma_{2}}{\gamma_{1}-\gamma_{2}}
  e^{-\frac{m\eta}{\hbar^{2}}L}e^{-\frac{m\eta}{\hbar^{2}}|x-L|}\right\} .
\end{equation}

Expressions for higher order terms of our perturbation theory can
be found also. Here we shall concentrate on identifying the second
order terms, since even higher order terms can be found by a
rather obvious extension of this procedure. For the purpose, we
had better rewrite the contributions in the right hand side of
(\ref{1st-int-eq}) appropriately. As regards the first term, we
may here use (instead of (\ref{it-rel})) the identity
\begin{equation}\label{it-rel-1}
  \hat{Q}_{k}=\hat{O}\hat{G}_{1}'+(\hat{V}_{2}-\delta
  E_{k})\hat{G}_{1}'\ ,\ \
  \left(\hat{G}_{1}'\equiv\sum_{n(\neq k)}\frac{|n\rangle\langle n|}
  {\varepsilon_{k}-\varepsilon_{n}}\right)
\end{equation}
to have it rewritten as
\begin{equation}\label{2nd-it-1term}
  \hat{Q}_{k}\hat{V}_{1}\hat{G}_{2}\hat{V}_{2}|k\rangle=
  \hat{Q}_{k}\hat{O}\hat{G}_{1}'\hat{V}_{1}\hat{G}_{2}\hat{V}_{2}|k\rangle+
  \hat{Q}_{k}(\hat{V}_{2}-\delta E_{k})\hat{G}_{1}'
  \hat{V}_{1}\hat{G}_{2}\hat{V}_{2}|k\rangle.
\end{equation}
On the other hand, with the second term, the relation
(\ref{it-rel}) can be used to write it as
\begin{equation}\label{2nd-it-2term}
  -\delta E_{k}\hat{Q}_{k}\hat{G}_{2}\hat{V}_{2}|k\rangle=
  -\delta E_{k}\hat{Q}_{k}\hat{O}(\hat{G}_{2})^{2}\hat{V}_{2}|k\rangle
  -\delta E_{k}\hat{Q}_{k}(\hat{V}_{1}-\delta E_{k})
  (\hat{G}_{2})^{2}\hat{V}_{2}|k\rangle.
\end{equation}
Using these forms with (\ref{1st-int-eq}) leads, after some
rearrangements, to the following equation:
\begin{equation}\label{2nd-it}\begin{array}{c}
  \hat{Q}_{k}\hat{O}\left\{|\delta\phi_{k}\rangle
  -|\delta\phi_{k}\rangle^{(1)}-
  \hat{G}_{1}'\hat{V}_{1}\hat{G}_{2}\hat{V}_{2}|k\rangle+
  \delta E_{k}(\hat{G}_{2})^{2}\hat{V}_{2}|k\rangle\right\}=\hspace{3cm}\\
  \hspace{3cm}\hat{Q}_{k}(\hat{V}_{2}-\delta E_{k})\hat{G}_{1}'
  \hat{V}_{1}\hat{G}_{2}\hat{V}_{2}|k\rangle-
  \delta E_{k}\hat{Q}_{k}(\hat{V}_{1}-\delta E_{k})
  (\hat{G}_{2})^{2}\hat{V}_{2}|k\rangle.
\end{array}\end{equation}

In the Appendix the expressions in the right hand side of
(\ref{2nd-it}) will be shown to be of higher order than the terms
appearing inside the curly bracket in the left hand side of the
same equation. Note that, for this behavior, it is crucial to have
$\hat{Q}_{k}\hat{V}_{1}\hat{G}_{2}\hat{V}_{2}|k\rangle$ rewritten
as in (\ref{2nd-it-1term}) and not by the form
\begin{equation}\label{2nd-it-1term-wrong}
  \hat{Q}_{k}\hat{V}_{1}\hat{G}_{2}\hat{V}_{2}|k\rangle=
  \hat{Q}_{k}\hat{O}\hat{G}_{2}\hat{V}_{1}\hat{G}_{2}\hat{V}_{2}|k\rangle+
  \hat{Q}_{k}(\hat{V}_{1}-\delta E_{k})\hat{G}_{2}
  \hat{V}_{1}\hat{G}_{2}\hat{V}_{2}|k\rangle,
\end{equation}
as the use of (\ref{it-rel}) would result in. [Here the term
involving $\hat{V}_{1}\hat{G}_{2}\hat{V}_{1}$ is dangerous, when
continuum contributions are considered.] On the other hand, as for
the term $-\delta
E_{k}\hat{Q}_{k}\hat{G}_{2}\hat{V}_{2}|k\rangle$, it is allowed to
have (\ref{2nd-it-2term}) replaced by another relation obtained
with the use of (\ref{it-rel-1}) --- but, using
(\ref{2nd-it-2term}) (and hence the equation (\ref{2nd-it})) leads
to a simpler perturbation theory \textit{practically}. Now, based
on this order count for the terms appearing on both sides of
(\ref{2nd-it}), we are led to conclude that the expression inside
the curly bracket may be set to zero in our present approximation.
Note that this reasoning is entirely similar to what we used with
(\ref{1st-int-eq}). As a result, it is found that our second order
approximation to $|\delta\phi_{k}\rangle$ can be identified with
\begin{equation}\label{2nd-eigen}
  |\delta\phi_{k}\rangle^{(2)}=
  \hat{G}_{1}'\hat{V}_{1}\hat{G}_{2}\hat{V}_{2}|k\rangle
  -\delta
  E_{k}^{(1)}\hat{G}_{2}\hat{G}_{2}\hat{V}_{2}|k\rangle.
\end{equation}
By using this expression with (\ref{gen-shift-formula}), one can
obtain the corresponding formula for the second-order energy shift
also.

For the third or higher order approximation, one may repeat the
above procedure. Clearly, the approximation at desired order
follows immediately once one has the appropriate generalization of
the equation like (\ref{1st-int-eq}) or (\ref{2nd-it}). As we have
explained above, such generalization can always be found by using
the identities (\ref{it-rel}) and (\ref{it-rel-1}) in a judicious
way with the corresponding equation one order lower. For a useful
guideline here, see the Appendix.

\section{(almost-) degenerate perturbation theory}

Our perturbation theory in the previous section was developed
under the no degeneracy assumption; that is, for a given
unperturbed state $|k\rangle$ (an eigenstate of $\hat{H}_{1}$ with
eigenvalue $\varepsilon_{k}$), no other eigenstate of
$\hat{H}_{1}$ or $\hat{H}_{2}$ has the corresponding eigenvalue
equal or very close to $\varepsilon_{k}$. In this section we will
dispense with this restrictive assumption. The perturbation theory
to be developed below is applicable to the case when (almost-)
degeneracy, within the spectrum of $\hat{H}_{1}$ or between the
spectra of the two local Hamiltonians $\hat{H}_{1}$ and
$\hat{H}_{2}$, is present. This consideration is especially
relevant since many physically interesting problems, which were
treated traditionally by the molecular orbital theory, do come
with such (almost-) degeneracy due to symmetry or by other
reasons.

First, we focus on the case when there are two (almost-)
degenerate states $|k\rangle$ and $|\bar{k}\rangle$ with
$\varepsilon_{k}\approx u_{\bar{k}}$. That is, each local
Hamiltonian has a bound state of almost identical energy. [ This
happens especially if $\hat{H}_{2}$ is related to $\hat{H}_{1}$ by
a simple spatial translation, i.e., $V_{2}(x)=V_{1}(x-L)$. ] In
this case we expect that the exact eigenstate $|\phi_{k}\rangle$
of the full Hamiltonian (\ref{tot-ham}) have large overlap with
both $|k\rangle$ and $|\bar{k}\rangle$, in accordance with the
philosophy of tight-binding approximation. So we may set up our
perturbation theory by writing
\begin{equation}\label{deg-ansatz}
  |\phi_{k}\rangle = |k\rangle+b|\bar{k}\rangle+|\delta\phi_{k}\rangle
\end{equation}
where $|\delta\phi_{k}\rangle$ is supposed to be small, but the
constant $b$ can be a priori ${\mathcal{O}}(1)$. [ Of course, if
$\varepsilon_{k}$ and $u_{k}$ were not very close to each other,
$b$ would become much smaller than $1$. ] Inserting the form
(\ref{deg-ansatz}) into the Schr\"{o}dinger equation
(\ref{original-sch}) then yields
\begin{equation}\label{deg-sch}
  \hat{O}|\delta\phi_{k}\rangle =
  (\hat{V}_{2}-\delta E_{k})|k\rangle +
  b(\hat{V}_{1}+u_{\bar{k}}-\varepsilon_{k}-\delta E_{k})|\bar{k}\rangle,
\end{equation}
where $\delta E_{k}=E_{k}-\varepsilon_{k}$, and $\hat{O}$ is the
operator introduced in (\ref{exact-prop}). Here the unknowns are
$\delta E_{k}$, $b$ and $|\delta\phi_{k}\rangle$, and
(\ref{deg-sch}) contains all the conditions required of them.

If we multiply both sides of (\ref{deg-sch}) by $\langle k|$ or
$\langle \bar{k}|$ on the left, we obtain two relations which can
be used to determine $\delta E_{k}$ and the constant $b$, given
the knowledge on $|\delta\phi_{k}\rangle$. Explicitly, we may
write them as two different expressions for $\delta E_{k}$, i.e.,
\begin{eqnarray}
  \delta E_{k}&=&\frac{\alpha+b\Gamma+
  \langle k|\hat{V}_{2}|\delta\phi_{k}\rangle}
  {1+b\Delta+\langle k|\delta\phi_{k}\rangle},\label{deg-energy-1}\\
  \delta E_{k}&=&\frac{b(u_{\bar{k}}-\varepsilon_{k})+\Gamma+b\beta+
  (u_{\bar{k}}-\varepsilon_{k})\langle \bar{k}|\delta\phi_{k}\rangle+
  \langle \bar{k}|\hat{V}_{1}|\delta\phi_{k}\rangle}
  {b+\Delta+\langle \bar{k}|\delta\phi_{k}\rangle},\label{deg-energy-2}
\end{eqnarray}
where $\alpha$, $\beta$, $\Gamma$ and $\Delta$ represent the
matrix elements
\begin{equation}\label{deg-elements}\begin{array}{ll}
\alpha\equiv\langle k|\hat{V}_{2}|k\rangle,&
\ \ \ \beta\equiv\langle\bar{k}|\hat{V}_{1}|\bar{k}\rangle,\\
\Gamma\equiv\langle k|\hat{V}_{2}|\bar{k}\rangle,& \ \ \
\Delta\equiv\langle k|\bar{k}\rangle.
\end{array}\end{equation}
[We have used the fact that, when $|k\rangle$ ($|\bar{k}\rangle$)
is a \textit{nondegenerate} eigenstate of $\hat{H}_{1}$
($\hat{H}_{2}$), it is possible to take $\alpha$, $\beta$,
$\Gamma$ and $\Delta$ to be real]. To fix the constant $b$ (for
given $|\delta\phi_{k}\rangle$), one can thus solve the quadratic
equation obtained by equating the two expressions in the right
hand sides of (\ref{deg-energy-1}) and (\ref{deg-energy-2}). Then,
how can one determine the eigenfunction correction
$|\delta\phi_{k}\rangle$? As in the nondegenerate case considered
in Sec.2, an appropriate perturbation theory for
$|\delta\phi_{k}\rangle$ may be set up by considering the
restriction imposed by (\ref{deg-sch}) on its components belonging
to the space orthogonal to $|k\rangle$ or $|\bar{k}\rangle$.
Again, in the corresponding development, we will not impose any
specific condition on $\langle k|\delta\phi_{k}\rangle$ (or, if
one wishes, on $\langle
k|\delta\phi_{k}\rangle+b\langle\bar{k}|\delta\phi_{k}\rangle$);
following the order-by-order analysis, it should suffice for us to
choose $|\delta\phi_{k}\rangle$ to be a simplest available
expression that is consistent with the equation (\ref{deg-sch}).

When the separation distance between the local potentials is taken
to be large, the leading approximation in our approach corresponds
to the standard molecular orbital theory. This can be seen as
follows. In (\ref{deg-energy-1}) and (\ref{deg-energy-2}), b is
order 1 while $|\delta\phi_{k}\rangle$ is supposed to contain at
least one wave-function stretching factor. Also, in the limit we
are considering, all four matrix elements in (\ref{deg-elements})
should be quite small; $\alpha$ and $\beta$ contain two
wave-function stretching factors, and $\Gamma$ and $\Delta$ one
wave-function stretching factor each. In view of the potential
$\hat{V}_{2}$ present in its definition, $\Gamma$ may be estimated
to be of order $|u_{\bar{k}}|\Delta$. Moreover, from the assumed
almost-degeneracy of the two states, it should be natural to
assume that
\begin{equation}\label{almost-deg}
  |u_{\bar{k}}-\varepsilon_{k}|\ll |u_{\bar{k}}|.
\end{equation}
With $\Gamma\sim |u_{\bar{k}}|\Delta$, this implies
$u_{\bar{k}}-\varepsilon_{k}\ll \Gamma/\Delta$ also. Then, in the
leading approxiamtion, we may set $|\delta\phi_{k}\rangle^{(0)}=0$
(i.e., $|\phi_{k}\rangle=|k\rangle+b|\bar{k}\rangle$ to this
order) and replace the right hand sides of (\ref{deg-energy-1})
and (\ref{deg-energy-2}) by  $b\Gamma$ and
$\frac{b(u_{\bar{k}}-\varepsilon_{k})+\Gamma}{b}$, respectively.
From these, we conclude that
\begin{equation}\label{near-deg-mix}
  b^{(0)}=\frac{u_{\bar{k}}-\varepsilon_{k}}{2\Gamma}\pm
  \sqrt{\left(\frac{u_{\bar{k}}-\varepsilon_{k}}{2\Gamma}\right)^{2}+1}
  \ \ ,\ \ \delta E^{(1)}=\Gamma b^{(0)}.
\end{equation}
These are what one would expect with the original Hamiltonian
replaced by the $2\times 2$ matrix Hamiltonian (in the space
spanned by two \textit{atomic orbitals} $|k\rangle$ and
$|\bar{k}\rangle$)
\begin{equation}\label{trunc-ham}
  \left(\begin{array}{cc}
    \varepsilon_{k}+\alpha&\Gamma\\
    \Gamma&u_{\bar{k}}+\beta
  \end{array}\right),
\end{equation}
and with $\alpha$, $\beta$ ignored because they contain two
wave-function stretching factors while $\Gamma$ has one. In
particular, if $u_{\bar{k}}-\varepsilon_{k}\ll \Gamma$, that is,
if two energies are very close, the expressions in
(\ref{near-deg-mix}) tend to the familiar values in the exactly
degenerate case, $b^{(0)}\sim\pm 1$ (i.e.,
$|\phi_{k}\rangle\sim|k\rangle\pm|\bar{k}\rangle$) and $\delta
E^{(1)}_{k} = \pm\Gamma$. It may also be of interest to look at
the case $u_{\bar{k}}-\varepsilon_{k}\gg \Gamma$, that is, when
the two energy values are not very close to each other (although
they are almost degenerate in the sense of (\ref{almost-deg})).
Then, from the two values given for $b^{(0)}$, only one of them
--- that with the behavior $b^{(0)}\rightarrow 0$ as $\Gamma$
approaches zero --- may be chosen since we are seeking for a
solution that reduces to $|k\rangle$ in the absence of the
potential $V_{2}$. Hence, with $u_{\bar{k}}-\varepsilon_{k}\gg
\Gamma$, we find from (\ref{near-deg-mix}) the values
$b^{(0)}=\frac{\Gamma}{\varepsilon_{k}-u_{\bar{k}}}$ and $\delta
E^{(1)}_{k}=\frac{\Gamma^{2}}{\varepsilon_{k}-u_{\bar{k}}}$, which
are the results we can infer also on the basis of our formulas
(\ref{low-state-exp}) and (\ref{low-shift-exp}) (i.e., the
lowest-order results in our nondegenerate formalism).

For higher order corrections, one should look for an iterative
solution of (\ref{deg-sch}), as we did the same with
(\ref{sch-decomp}) in the nondegenerate case. Here, for successive
iteration, we will make use of the relation (instead of
(\ref{it-rel}))
\begin{equation}\label{it-rel-2}
  \hat{Q}_{\bar{k}}=\hat{O}\hat{G}_{2}'+(\hat{V}_{1}-\delta
  E_{k})\hat{G}_{2}'\ ,
\end{equation}
where $\hat{Q}_{\bar{k}}\equiv 1 -|\bar{k}\rangle\langle\bar{k}|$,
and
\begin{equation}\label{free-prop-2}
 \hat{G}_{2}'\equiv\hat{Q}_{\bar{k}}\hat{G}_{2}\hat{Q}_{\bar{k}}
 =\sum_{\bar{n}(\neq\bar{k})}\frac{|\bar{n}\rangle\langle\bar{n}|}
 {\varepsilon_{k}-u_{\bar{n}}}\ .
\end{equation}
The Green's operator $\hat{G}_{1}'$, satisfying (\ref{it-rel-1}),
will be useful as well. But we will here proceed somewhat
differently from the nondegenerate case by not utilizing a
suitably projected version of (\ref{deg-sch}) in making iteration;
for the present (almost-)degenerate case, manipulating directly
with (\ref{deg-sch}) is more convenient. Now note that, thanks to
(\ref{it-rel-1}) and (\ref{it-rel-2}), the terms
$\hat{V}_{2}|k\rangle$ and $\hat{V}_{1}|\bar{k}\rangle$ in
(\ref{deg-sch}) can be rewritten as
\begin{eqnarray}
  \hat{V}_{2}|k\rangle&=&\hat{Q}_{\bar{k}}\hat{V}_{2}|k\rangle+
  |\bar{k}\rangle\langle\bar{k}|\hat{V}_{2}|k\rangle\nonumber\\
  &=&\hat{O}\hat{G}_{2}'\hat{V}_{2}|k\rangle+
  (\hat{V}_{1}-\delta E_{k})\hat{G}_{2}'\hat{V}_{2}|k\rangle+
  \Gamma|\bar{k}\rangle,\label{deg-it-1}\\
  \hat{V}_{1}|\bar{k}\rangle&=&\hat{Q}_{k}\hat{V}_{1}|\bar{k}\rangle+
  |k\rangle\langle k|\hat{V}_{1}|\bar{k}\rangle\nonumber\\
  &=&\hat{O}\hat{G}_{1}'\hat{V}_{1}|\bar{k}\rangle+
  (\hat{V}_{2}-\delta E_{k})\hat{G}_{1}'\hat{V}_{1}|\bar{k}\rangle+
  \{\Gamma-(u_{\bar{k}}-\varepsilon_{k})\Delta\}|k\rangle.\label{deg-it-2}
\end{eqnarray}
Using these in (\ref{deg-sch}) and then collecting all terms
involving the operator $\hat{O}$ explicitly, we obtain the
following equation:
\begin{equation}\label{deg-sch-2}\begin{array}{l}
  \hspace{-1cm}\hat{O}\left(|\delta\phi_{k}\rangle
  -\hat{G}_{2}'\hat{V}_{2}|k\rangle-
  b\hat{G}_{1}'\hat{V}_{1}|\bar{k}\rangle\right)=
  \hat{V}_{1}\hat{G}_{2}'\hat{V}_{2}|k\rangle-
  \delta E_{k}\hat{G}_{2}'\hat{V}_{2}|k\rangle+
  b\hat{V}_{2}\hat{G}_{1}'\hat{V}_{1}|\bar{k}\rangle\\
  \hspace*{2cm}-
  b\delta E_{k}\hat{G}_{1}'\hat{V}_{1}|\bar{k}\rangle
  -\{\delta E_{k}-b\Gamma\left(1-\frac{u_{\bar{k}}\!-\!
  \varepsilon_{k}}{\Gamma/\Delta}\right)\}|k\rangle-
  \{b\delta E_{k}-\Gamma\}|\bar{k}\rangle.
\end{array}\end{equation}

Based on (\ref{deg-sch-2}), we will now show that the leading
approximation for $|\delta\phi_{k}\rangle$ can be taken as
\begin{equation}\label{deg-1st-eigen}
  |\delta\phi_{k}\rangle^{(1)}=
  \hat{G}_{2}'\hat{V}_{2}|k\rangle+b\hat{G}_{1}'\hat{V}_{1}|\bar{k}\rangle.
\end{equation}
First note that, as in the nondegenerate case, the first four
terms in the right hand side of (\ref{deg-sch-2}) can be shown to
be of higher order than the expression
$\hat{G}_{2}'\hat{V}_{2}|k\rangle+b\hat{G}_{1}'\hat{V}_{1}|k\rangle$.
On the other hand, the last two terms in the right hand side of
(\ref{deg-sch-2}) are explicitly proportional to $|k\rangle$ or
$|\bar{k}\rangle$; they are present because we are not working
with a projected equation.Still, we observe that these terms are
also smaller than the expression
$\hat{G}_{2}'\hat{V}_{2}|k\rangle+b\hat{G}_{1}'\hat{V}_{1}|k\rangle$,
if the lowest order values for $\delta E_{k}$ and $b$ (in
(\ref{near-deg-mix})) are used. [Here remember that
$\frac{u_{\bar{k}}\!-\!\varepsilon_{k}}{\Gamma/\Delta}\ll 1$.]
Then, based on these and our earlier observation as regards the
effect of the operator $\hat{O}$ (in that case with
(\ref{1st-int-eq})), the identification (\ref{deg-1st-eigen}) can
be made.

The expression (\ref{deg-1st-eigen}), with $b$ replaced by
$b^{(0)}$, may in turn be used in (\ref{deg-energy-1}) and
(\ref{deg-energy-2}) to find the second order energy shift $\delta
E_{k}^{(2)}$ and the value $b^{(1)}$. The results, to the
appropriate order in the wave-function stretching factor (but
without making an expansion with respect to another small factor
$\frac{|u_{\bar{k}}-\varepsilon_{k}|}{\Gamma/\Delta}$), read
\begin{eqnarray}
  b^{(1)}\ &=&\left\{\frac{\beta-\alpha}{2\Gamma}+
  \frac{\langle\bar{k}|\hat{V}_{1}\hat{G}_{1}'\hat{V}_{1}|\bar{k}\rangle
  -\langle k|\hat{V}_{2}\hat{G}_{2}'\hat{V}_{2}|k\rangle}{2\Gamma}\right\}
  \left\{1\pm\frac{\frac{u_{\bar{k}}-\varepsilon_{k}}{2\Gamma}}
  {\sqrt{1+\left(\frac{u_{\bar{k}}-\varepsilon_{k}}
  {2\Gamma}\right)^{2}}}\right\},\label{almost-deg-2nd-coeff}\\
  \delta E_{k}^{(2)}&=&\Gamma b^{(1)}+\alpha+
  \langle k|\hat{V}_{2}\hat{G}_{2}'\hat{V}_{2}|k\rangle-
  \Gamma\Delta(b^{(0)})^{2}.\label{almost-deg-2nd-energy}
\end{eqnarray}
Especially, with $u_{\bar{k}}=\varepsilon_{k}$, i.e., in exactly
degenerate case, (\ref{almost-deg-2nd-energy}) reduces to
\begin{equation}\label{deg-2nd-energy}
  \delta E_{k}^{(2)}\!=\!\frac{\langle k|\hat{V}_{2}|k\rangle\!+\!
  \langle k|\hat{V}_{2}\hat{G}_{2}'\hat{V}_{2}|k\rangle}{2}\!+\!
  \frac{\langle\bar{k}|\hat{V}_{1}|\bar{k}\rangle\!+\!
  \langle\bar{k}|\hat{V}_{1}\hat{G}_{1}'\hat{V}_{1}|\bar{k}\rangle}{2}\!-\!
  \langle k|\bar{k}\!\rangle\langle k|\hat{V}_{2}|\bar{k}\rangle,
\end{equation}
as the definitions for $\alpha$, $\beta$, $\Gamma$ and $\Delta$ in
(\ref{deg-elements}) are used. According to this formula, the
second order energy shifts for the two split states become
identical. An explicit check for the validity of
(\ref{deg-2nd-energy}) may be made for our $\delta$-function
example (see (\ref{delta-pot})) with
$\gamma_{1}=\gamma_{2}\equiv\gamma$. According to the direct
calculation based on (\ref{delta-energy-eq}), we have $\delta
E_{k}^{(2)}=-\frac{m\gamma^{2}}{2\hbar^{2}} \left(2\frac{m\gamma
L}{\hbar^{2}}+1\right)e^{-2\frac{m\gamma L}{\hbar^{2}}}$. We have
verified that this very result is reproduced when various terms in
(\ref{deg-2nd-energy}) are explicitly evaluated. Also, as in the
nondegenerate case, a further rearrangement of (\ref{deg-sch-2})
may be considered to obtain the expressions for the next order
contributions. But, because of the complications involved and
because their usefulness is rather limited, we will not consider
such further higher order terms.

It is possible to generalize the above discussion to the case when
there are more than two degenerate states, that is, $N_{1}$
eigenstates $\{|k_{\mu}\rangle:\mu=1,\cdots,N_{1}\}$ of
$\hat{H}_{1}$ with the given energy $\varepsilon_{k}$ and $N_{2}$
eigenstates
$\{|\bar{k}_{\bar{\mu}}\rangle:\bar{\mu}=1,\cdots,N_{2}\}$ of
$\hat{H}_{2}$, with the same energy $u_{\bar{k}}=\varepsilon_{k}$.
Here we will concentrate on exactly degenerate case, not to make
the problem too complicated. Now, for the exact eigenstates of the
total Hamiltonian, we may write
\begin{equation}
  |\phi\rangle= \sum_{\mu}a_{\mu}|k_{\mu}\rangle+
  \sum_{\bar{\mu}}b_{\bar{\mu}}|\bar{k}_{\bar{\mu}}\rangle+
  |\delta\phi\rangle,
\end{equation}
where $a_{\mu}$ and $b_{\bar{\mu}}$ can be ${\mathcal{O}}(1)$, but
$|\delta\phi\rangle$ is small. Inserting this form into the
Schr\"{o}dinger equation (\ref{original-sch}), we obtain an
equation similar to (\ref{deg-sch}),
\begin{equation}\label{multi-deg-sch}
  \hat{O}|\delta\phi\rangle =
  \sum_{\mu}a_{\mu}(\hat{V}_{2}-\delta E_{k})|k_{\mu}\rangle +
  \sum_{\bar{\mu}}b_{\bar{\mu}}(\hat{V}_{1}-\delta E_{k})
  |\bar{k}_{\bar{\mu}}\rangle.
\end{equation}
Then, from multiplying both sides of this equation by $\langle
k_{\mu}|$ and $\langle\bar{k}_{\bar{\mu}}|$ from the left, we
obtain the following conditions which may be used to determine
$\delta E$, $a_{\mu}$ and $b_{\bar{\mu}}$:
\begin{eqnarray}
  &&\delta E(a_{\mu}+\sum_{\bar{\nu}}\Delta_{\mu\bar{\nu}}b_{\bar{\nu}}+
  \langle k_{\mu}|\delta\phi\rangle)=
  \sum_{\bar{\nu}}\Gamma_{\mu\bar{\nu}}b_{\bar{\nu}}+
  \sum_{\nu}\alpha_{\mu\nu}a_{\nu}+\langle k_{\mu}|\hat{V}_{2}
  |\delta\phi\rangle,\label{multi-deg-1}\\
  &&\delta E (b_{\bar{\mu}}+\sum_{\nu}\Delta^{\dag}_{\bar{\mu}\nu}a_{\nu}+
  \langle\bar{k}_{\bar{\mu}}|\delta\phi\rangle)=
  \sum_{\nu}\Gamma^{\dag}_{\bar{\mu}\nu}a_{\nu}+
  \sum_{\bar{\nu}}\beta_{\bar{\mu}\bar{\nu}}b_{\bar{\nu}}+\langle\bar{k}_{\bar{\mu}}|\hat{V}_{1}
  |\delta\phi\rangle,\label{multi-deg-2}
\end{eqnarray}
where we have defined
\begin{equation}\label{deg-elements-multi}\begin{array}{ll}
\alpha_{\mu\nu}\equiv\langle k_{\mu}|\hat{V}_{2}|k_{\nu}\rangle,&\
\beta_{\bar{\mu}\bar{\nu}}\equiv\langle
\bar{k}_{\bar{\mu}}|\hat{V}_{1}|\bar{k}_{\bar{\nu}}\rangle,\\
\Gamma_{\mu\bar{\nu}}\equiv\langle
k_{\mu}|\hat{V}_{1}|\bar{k}_{\bar{\nu}}\rangle=\langle
k_{\mu}|\hat{V}_{2}|\bar{k}_{\bar{\nu}}\rangle,&\
\Delta_{\mu\bar{\nu}}\equiv \langle
k_{\mu}|\bar{k}_{\bar{\nu}}\rangle,
\end{array}\end{equation}
and $\dag$ denotes the hermitian conjugate. [Note that $\Gamma$
and $\Delta$ are $N_{1}\times N_{2}$ matrices --- not square
matrices in general.]

To determine the lowest order values $\delta E^{(1)}$,
$a_{\mu}^{(0)}$ and $b_{\bar{\mu}}^{(0)}$, we note that
(\ref{multi-deg-1}) and (\ref{multi-deg-2}), as only leading order
terms are kept, imply the following equations:
\begin{equation}\label{multi-1st-eigen}\begin{array}{l}
  \displaystyle\delta E^{(1)}a_{\mu}^{(0)}=
  \sum_{\bar{\nu}}\Gamma_{\mu\bar{\nu}}b_{\bar{\nu}}^{(0)}\
  ,\ \ (\mu=1,\cdots,N_{1})\\
  \displaystyle\delta E^{(1)}b_{\bar{\mu}}^{(0)}=
  \sum_{\nu}\Gamma^{\dag}_{\bar{\mu}\nu}a_{\nu}^{(0)}\ ,
  \ \ (\bar{\mu}=1,\cdots,N_{2}).
\end{array}\end{equation}
These can be regarded as a single eigenvector equation for an
$(N_{1}\!+\!N_{2})$-vector $(a^{(0)},b^{(0)})$,
\begin{equation}\label{large-matrix}
  \left(\begin{array}{cc}0\ &\Gamma\\ \Gamma^{\dag}&0\end{array}\right)
  \left(\begin{array}{c}a^{(0)}\\b^{(0)}\end{array}\right)=
  \delta
  E^{(1)}\left(\begin{array}{c}a^{(0)}\\b^{(0)}\end{array}\right).
\end{equation}
This is equivalent to the molecular orbital theory approximation
[1,2] in which the full Hilbert space is truncated to the
finite-dimensinal space spanned by $N_{1}\!+\!N_{2}$ atomic
orbitals, i.e., $\{|k_{\mu}\rangle\}$ and
$\{|\bar{k}_{\bar{\mu}}\rangle\}$. Assuming $N_{1}\geq N_{2}$, the
$N_{1}\!+\!N_{2}$ eigenvectors and corresponding eigenvalues may
schematically be expressed by the forms
\begin{equation}\label{multi-eigen}\begin{array}{l}
  (a^{(0)}_{I\mu},b^{(0)}_{I\mu})=(u_{I\mu},\pm v_{I\mu}),\ \ \
  \delta E^{(1)}_{I}=\pm\lambda_{I},\ \ {\rm for}\ I=1,2,\cdots N_{2}\\
  (a^{(0)}_{J\mu},b^{(0)}_{J\mu})=(U_{J\mu},0),\ \ \
  \delta E^{(1)}_{J}=0,\ \ {\rm for}\  J=2N_{2}+1,\cdots N_{1}\!+\!N_{2}.
\end{array}\end{equation}
The first $2N_{2}$ eigenvectors are given by $N_{2}$ pairs of
states, i.e., $\{\displaystyle\sum_{\mu}u_{I
\mu}|k_{\mu}\rangle\pm
\sum_{\bar{\mu}}v_{I\bar{\mu}}|\bar{k}_{\bar{\mu}}\rangle\ ;
I=1,\cdots,N_{2}\}$, with respective energy splits
$\pm\lambda_{I}$. [Here, from studying (\ref{large-matrix}), it
can be shown that
$\displaystyle\sum_{\mu}u_{I\mu}^{\ast}u_{I\mu}=\sum_{\mu}v_{I\mu}^{\ast}v_{I\mu}$
for each $I=1,2,\cdots, N_{2}$, and so all $u_{I}$, $v_{I}$ may be
taken to be unit vectors.] If the eigenvalue set
$\{\pm\lambda^{i}\}$ contains zero or the same value more than
once, the degeneracy is not completely lifted and one may have to
perform higher order analysis for the effect (and associated true
energy eigenvectors). There is no first-order energy shift for the
remaining $N_{1}\!-\!N_{2}$ eigenstates, represented by
$\{\displaystyle\sum_{\mu}U_{J \mu}|k_{\mu}\rangle\ ;
J=2N_{2}+1,\cdots,N_{1}+N_{2}\}$. Hence, with $N_{1}>N_{2}$, there
always remains some energy degeneracy which is not lifted by the
lowest order consideration alone.

To develop the corresponding higher-order perturbation theory, one
should now take the expressions
\begin{equation}
  \hat{G}_{1}'\equiv\sum_{n(\neq k_{\mu})}
  \frac{|n\rangle\langle n|}{\varepsilon_{k}-\varepsilon_{n}}\ \ ,
  \ \ \hat{G}_{2}'\equiv\sum_{\bar{n}(\neq\bar{k}_{\bar{\mu}})}
  \frac{|\bar{n}\rangle\langle\bar{n}|}{\varepsilon_{k}-u_{\bar{n}}}
\end{equation}
as relevant Green's functions and proceed in more or less the same
manner as in our earlier consideration. Especially, with
$N_{1}=N_{2}\equiv N$, we then find the results (as direct
generalizations of (\ref{deg-1st-eigen}) and
(\ref{deg-2nd-energy}))
\begin{eqnarray}
  |\delta\phi_{I}\rangle^{(1)}&=&\sum_{\mu}u_{I\mu}\hat{G}_{2}'
  \hat{V}_{2}|k_{\mu}\rangle\pm\sum_{\bar{\mu}}
  v_{I\bar{\mu}}\hat{G}_{1}'\hat{V}_{1}|\bar{k}_{\bar{\mu}}\rangle,\label{equal-multi-2nd-eigen}\\
  \delta E_{I}^{(2)}\ &=&\frac{1}{2}\sum_{\mu,\nu}u^{\ast}_{I\mu}\left\{\alpha_{\mu\nu}
  +\langle k_{\mu}|\hat{V}_{2}\hat{G}_{2}'\hat{V}_{2}|k_{\nu}\rangle\right\}u_{I\nu}+
  \frac{1}{2}\sum_{\bar{\mu},\bar{\nu}}v^{\ast}_{I\bar{\mu}}\left\{\beta_{\bar{\mu}\bar{\nu}}
  +\langle \bar{k}_{\bar{\mu}}|\hat{V}_{1}\hat{G}_{1}'\hat{V}_{1}|\bar{k}_{\bar{\nu}}\rangle
  \right\}v_{I\bar{\nu}}\nonumber\\
  &&\hspace{-0.2cm}-\frac{\lambda_{I}}{2}\sum_{\mu,\bar{\nu}}\left\{u^{\ast}_{I\mu}
  \Delta_{\mu\bar{\nu}}v_{I\bar{\nu}}+v^{\ast}_{I\bar{\nu}}\Delta^{\dag}_{\bar{\nu}\mu}
  u_{I\mu}\right\},\label{equal-multi-2nd-energy}
\end{eqnarray}
where we have normalized $u_{I}$ and $v_{I}$ to be unit vectors.

With $N_{1}>N_{2}$, we need to consider also the higher order
terms to determine the above $N_{1}\!-\!N_{2}$ eigenvectors
$\{\displaystyle\sum_{\mu}U_{I \mu}|k_{\mu}\rangle\ ;
J=2N_{2}+1,\cdots,N_{1}+N_{2}\}$ unambiguously and the possible
energy splitting between them. To that end, one has to study the
second order contributions from (\ref{multi-deg-1}). Let us here
assume for simplicity that the $(N_{1}\!-\!N_{2})$-dimensional
space spanned by the states $\displaystyle\sum_{\mu}U_{I
\mu}|k_{\mu}\rangle$ represent the \textit{entire} subspace with
$\delta E^{(1)}=0$ in the space of atomic orbitals. Then observe
that, in view of the second relation in (\ref{multi-1st-eigen}),
this $(N_{1}\!-\!N_{2})$-dimensional space with $\delta E^{(1)}=0$
can be identified with the kernel of the matrix $\Gamma^{\dag}$.
One now finds from (\ref{multi-deg-1}) that the (yet unknown)
coefficients $U_{J\mu}$ should be associated with the solutions of
\begin{equation}\label{multi-2nd-eigen}
  \delta E^{(2)}_{J}U_{J\mu}=\sum_{\nu}\alpha_{\mu\nu}U_{J\nu}+
  \sum_{\bar{\nu}}\Gamma_{\mu\bar{\nu}}b^{(1)}_{J\bar{\nu}}+
  \langle k_{\mu}|\hat{V}_{2}|\delta\phi_{J}\rangle^{(1)},
\end{equation}
for $|\delta\phi_{J}\rangle^{(1)}$ expressed in terms of
$U_{J\mu}$ through
$|\delta\phi_{J}\rangle^{(1)}=\sum_{\mu}U_{J\mu}\hat{G}_{2}'
\hat{V}_{2}|k_{\mu}\rangle$ (see (\ref{equal-multi-2nd-eigen})).
Actually, in (\ref{multi-2nd-eigen}), it can be shown (using the
property $\sum_{\nu}\Gamma^{\dag}_{\bar{\mu}\nu}U_{J\nu}=0$) that
the term $\displaystyle
\sum_{\bar{\nu}}\Gamma_{\mu\bar{\nu}}b^{(1)}_{J\bar{\nu}}$ is
irrelevant to this order, and therefore (\ref{multi-2nd-eigen}) is
really an eigenvector equation for the vectors $U_{J\mu}$:
\begin{equation}\label{unequal-multi}
  \delta E^{(2)}_{J}U_{J\mu}=\sum_{\nu}\left(\langle k_{\mu}|\hat{V}_{2}
  |k_{\nu}\rangle+\langle k_{\mu}|\hat{V}_{2}\hat{G}_{2}'\hat{V}_{2}|k_{\nu}\rangle
  \right)U_{J\nu}\ .
\end{equation}
This equation may be used to determine the coefficients $U_{J\mu}$
and the energy shifts $\delta E_{J}^{(2)}$. As one can see from
this consideration, our perturbative formalism can deal with
essentially all situations regarding the bound-state problem with
well-separated potentials.

\section{concluding remarks}

In this paper we have presented a systematic perturbation theory
for energy eigenstates when the potential of the system consists
of two spatially well-separated pieces, under the assumption that
complete energy eigenstates of the two local Hamiltonians are
available for our use. Our perturbative development, an expansion
in the number of wave-function stretching factors, is reminiscent
of the multiple scattering series. Depending on whether the local
Hamiltonians have (almost-)degenerate energy levels or not,
different perturbation theories must be used. Especially, when the
local Hamiltonians have degenerate energy levels, one obtains from
our theory systematic higher-order correction terms beyond the
predictions of the molecular orbital theory. The reasonably simple
formulas we found for the leading correction terms, that is,
(\ref{1st-state}) and (\ref{1st-shift}) in the nondegenerate case
and (\ref{deg-1st-eigen})-(\ref{deg-2nd-energy}) (or
(\ref{equal-multi-2nd-eigen})-(\ref{unequal-multi})) in the
degenerate case, may have some immediate practical applications.

Extension to the case with more than two spatially localized
potentials (in fact even to the case of a lattice of potentials)
should be straightforward. Also, if the degenerate atomic orbitals
are present in association with certain symmetry in the system,
one may utilize so-called symmetry-adapted linear combinations of
atomic orbitals[1,2] to simplify the perturbation theory. But we
have not made any systematic attempt in this direction. We also
remark that if the local potentials happen to be not sufficiently
well-localized (i.e., individual potentials have some long-range
tails), certain rearrangements may become necessary with our
perturbation series. This case deserves further study. One can
also contemplate on a simple field-theoretic application:
perturbation theory similar to the one given in this paper may be
used to study the fermionic bound states associated with a
soliton-antisoliton pair\cite{alt}.

\vskip 1cm
\parindent 0cm

{\bf\small ACKNOWLEDGEMENTS}

\vskip 0.4cm
\parindent 0.5cm

This work was supported in part by the BK21 project of the
Ministry  of Education, Korea, and the Korea Research Foundation
Grant 2001-015-DP0085.

\vskip 1cm
\parindent 0cm

{\bf\small APPENDIX}

\vskip 0.4cm
\parindent 0.5cm

In this appendix we will first present the argument that shows why
the terms in the right hand sides of (\ref{1st-int-eq}) and
(\ref{2nd-it}) are expected to be of higher order than those terms
in the left hand sides of the respective equations. We will then
make estimates, by general argument and by considering explicitly
the case of $\delta$-function potentials, on how small the
suppressed continuum contributions might be.

Let us start with our equation (\ref{1st-int-eq}), used for the
leading order approximation, and (\ref{1st-it-wrong}) for
comparison's sake. The two candidates one wishes to identify as
the leading-order expression of $|\delta\phi_{k}\rangle$, i.e.,
$\hat{G}_{2}\hat{V}_{2}|k\rangle$ according to (\ref{1st-int-eq})
and $\hat{G}_{1}'\hat{V}_{2}|k\rangle$ from (\ref{1st-it-wrong}),
contain one  wave-function stretching factor coming from the
overlap of $\hat{V}_{2}$ and $|k\rangle$. Since $\delta E_{k}$
carries at least two  wave-function stretching factors, the last
terms in the right hand sides of (\ref{1st-int-eq}) and
(\ref{1st-it-wrong}) can safely be ignored in lowest order
consideration. Here, the dangerous terms are the ones without
$\delta E_{k}$, i.e., $\hat{V}_{1}\hat{G}_{2}\hat{V}_{2}|k\rangle$
in (\ref{1st-int-eq}) and
$\hat{V}_{2}\hat{G}_{1}'\hat{V}_{2}|k\rangle$ in
(\ref{1st-it-wrong}). Both have clearly one  wave-function
stretching factor from $\hat{V}_{2}|k\rangle$, and one might
expect that an additional suppression might result from the
combination $\hat{V}_{1}\hat{G}_{2}\hat{V}_{2}$ or
$\hat{V}_{2}\hat{G}_{1}'\hat{V}_{2}$. The Green's operators
$\hat{G}_{1}'$ and $\hat{G}_{2}$ come with the sum over
appropriate energy eigenstates (of the local Hamiltonians
$\hat{H}_{1}$, $\hat{H}_{2}$), which include the continuum. As for
the bound state contributions of one local Hamiltonian to the
Green's operator, there should be such additional suppression (due
to small overlap) if they get combined with the potential of the
other local Hamiltonian. But, for the contribution to the Green's
operator from continuum states which are not localized at all, one
might not expect such suppression factor to show up, for these
continuum states would apparently have more or less equal overlap
regardless of the `location' of the other potential. But this
ignores the fact that one should really consider the net effect of
\textit{entire continuum states}. As will be discussed below, we
get a very different picture after integrating over the continuum.

Representing the continuous eigenstates of $\hat{H}_{2}$ by
$|\bar{q}\rangle$, we may express the continuum contribution of
$\hat{V}_{1}\hat{G}_{2}\hat{V}_{2}$ as
\begin{equation}\label{conti-contrib}
  \int d\bar{q}
  \frac{\hat{V}_{1}|\bar{q}\rangle\langle
  \bar{q}|\hat{V}_{2}}{\varepsilon_{k}-u_{\bar{q}}}\ \ .
\end{equation}
Here, for well-localized potentials $V_{1}$ and $V_{2}$, the
vectors $\hat{V}_{1}|\bar{q}\rangle\!=\!\int dx|x\rangle
V_{1}(x)\langle x|\bar{q}\rangle$ and
$\langle\bar{q}|\hat{V}_{2}\!=\!\int dy \langle\bar{q}|y\rangle
V_{2}(y)\langle y|$ will receive nonnegligible contributions
mainly from the regions around the respective potential centers,
i.e., $x=0$ and $y=L$. Furthermore, outside the range of the
potential $V_{2}$, the function $\langle x|\bar{q}\rangle$ may
well be approximated by a plane wave. This implies that
$\hat{V}_{1}|\bar{q}\rangle\langle\bar{q}|\hat{V}_{2}$ comes with
a phase factor $\langle x(\approx\!
0)|\hat{V}_{1}|\bar{q}\rangle\langle\bar{q}|\hat{V}_{2}|y(\approx\!
L)\rangle\sim e^{i\bar{q}L}$, which causes a destructive
interference if the separation $L$ is sufficiently large. Hence
the term in (\ref{conti-contrib}) comes with desired additional
suppression. On the other hand, an analogous consideration with
the continuum contribution of $\hat{V}_{2}\hat{G}_{1}'\hat{V}_{2}$
does not lead to such a fast oscillating factor and so no
suppression after summing over all corresponding continuum states.
This explains why, for our leading order analysis, we can utilize
(\ref{1st-int-eq}), but not (\ref{1st-it-wrong}).

By same reasoning as above, we expect that the continuum
contributions for, say, $\langle k|\hat{G}_{2}\hat{V}_{2}$ or
$\hat{V}_{2}(\hat{G}_{1}')^{n}\hat{V}_{1}$ (with $n\geq 2$) be
also suppressed. Suppression in the former case follows since
$\hat{G}_{2}$ appears between the bound state $|k\rangle$, which
is localized around the center of $V_{1}$, and the potential
$V_{2}$ (localized around $x=L$). As for the latter, the continuum
contributions from $(\hat{G}_{1}')^{n}$ give rise to a rapidly
oscillating phase if sandwiched between $\hat{V}_{1}$ and
$\hat{V}_{2}$. This suggests also a useful guideline in our
consideration of higher order perturbation terms: to have the
continuum contributions from $\hat{G}_{1}'$ and $\hat{G}_{2}$
suppressed as much as possible, we had better iterate the relevant
equation so that such Green's operator may take its place between
$\hat{V}_{1}$ (or $|k\rangle$) and $\hat{V}_{2}$. In fact, we
followed this guideline to obtain the expression for
$|\delta\phi\rangle^{(2)}$, i.e., when we proceeded from
(\ref{1st-int-eq}) to (\ref{2nd-it}). Of course, to confirm that
(\ref{2nd-it}) leads to the identification (\ref{2nd-eigen}) for
$|\delta\phi\rangle^{(2)}$, we need to pay more careful attention
to the order of various terms appearing in (\ref{2nd-it}), and
especially demonstrate the relative higher-order nature for the
expression on its right hand side. For this, see below.

First, with (\ref{2nd-it}), look at the terms appearing inside the
curly brackets on its left hand side. Based on (\ref{1st-state}),
we know that $|\delta\phi_{k}\rangle^{(1)}$ is of order
$e^{-\frac{1}{\hbar}\!\sqrt{2m|\varepsilon_{k}|}\ L}$; the
suppression factor here originates from the exponential tail of
the bound state $|k\rangle$. With the next term
$\hat{G}_{1}'\hat{V}_{1}\hat{G}_{2}\hat{V}_{2}|k\rangle$, we note
that (aside from $\hat{V}_{2}|k\rangle$, itself of order
$e^{-\frac{1}{\hbar}\!\sqrt{2m|\varepsilon_{k}|}\ L}$) there is
another small factor, say $\kappa_{2}$, coming from
$\hat{V}_{1}\hat{G}_{2}\hat{V}_{2}$. That is, this term is of
order $\kappa_{2}e^{-\frac{1}{\hbar}\!\sqrt{2m|\varepsilon_{k}|}\
L}$. The last term, $\delta
E_{k}(\hat{G}_{2})^{2}\hat{V}_{2}|k\rangle$, is of order
$\{e^{-\frac{1}{\hbar}\!\sqrt{2m|\varepsilon_{k}|}\ L}\}^{3}$.
Between $\hat{G}_{1}'\hat{V}_{1}\hat{G}_{2}\hat{V}_{2}|k\rangle$
and $\delta E_{k}(\hat{G}_{2})^{2}\hat{V}_{2}|k\rangle$ we can not
say generally which one is larger, because the magnitude of
$\kappa_{2}$ depends on the specific problem under study.
Therefore, it is appropriate to include \textit{both} terms in our
second order approximation, under the understanding that only one
term may well be dominant over the other in a given specific
problem. Similar analysis can also be made for various terms in
the right hand side of (\ref{2nd-it}). If we denote the small
factors emerging from $\hat{V}_{2}\hat{G}_{1}'\hat{V}_{1}$ and
$\hat{V}_{1}(\hat{G}_{2})^{2}\hat{V}_{2}$ by $\kappa_{1}$ and
$\kappa_{2}'$, respectively, we here find
\begin{eqnarray}
  &&\hat{V}_{2}\hat{G}_{1}'\hat{V}_{1}\hat{G}_{2}\hat{V}_{2}|k\rangle\sim\kappa_{1}\kappa_{2}
  e^{-\frac{1}{\hbar}\!\sqrt{2m|\varepsilon_{k}|}\ L}\ \ ,\ \delta
  E_{k}\hat{G}_{1}'\hat{V}_{1}\hat{G}_{2}\hat{V}_{2}|k\rangle\sim\kappa_{2}
  (e^{-\frac{1}{\hbar}\!\sqrt{2m|\varepsilon_{k}|}\ L})^{3}\ ,\nonumber\\
  &&\delta
  E_{k}\hat{V}_{1}(\hat{G}_{2})^{2}\hat{V}_{2}|k\rangle\sim
  \kappa_{2}'(e^{-\frac{1}{\hbar}\!\sqrt{2m|\varepsilon_{k}|}\
  L})^{3}\ \ ,\ (\delta
  E_{k})^{2}(\hat{G}_{2})^{2}\hat{V}_{2}|k\rangle\sim
  (e^{-\frac{1}{\hbar}\!\sqrt{2m|\varepsilon_{k}|}\ L})^{5}.
\end{eqnarray}
This shows that the terms in the right hand side of (\ref{2nd-it})
are relatively of higher order, as compared with those terms
appearing inside the curly brackets on its left hand side. [Note
that this is true without any extra assumption about the relative
ratios between $\kappa_{1}$, $\kappa_{2}$ and $\kappa_{2}'$].Hence
our formula for the second order correction
$|\delta\phi_{k}\rangle^{(2)}$ in (\ref{2nd-eigen}) follows.

In the above discussion, various \textit{small} factors like
$\kappa_{1}$, $\kappa_{2}$ have been introduced. How small are the
continuum contributions associated with these factors? Consider
$\hat{V}_{2}\hat{G}_{1}'\hat{V}_{1}$, with the related continuum
contribution given by
\begin{equation}\label{conti-app}
  \int dxdy \int dq V_{2}(x)\frac{\psi_{q}(x+L)\psi_{q}^{\ast}(y)}
  {-|\varepsilon_{k}|-\frac{\hbar^{2}q^{2}}{2m}}V_{1}(y),
\end{equation}
where $\psi_{q}(x)\equiv\langle x|q\rangle$, a continuous
eigenstate of $\hat{H}_{1}$. Here, for well-localized local
potentials, the integral will get most of its contribution from
the neighborhood of $x,y\sim 0$. For fixed $x$ in the
neighborhood, we may then extract the leading $L$-dependence from
$\psi_{q}(x+L)$ as
\begin{equation}\label{scatt}
  \psi_{q}(x)\approx A(q)e^{i\Delta(k)}e^{iq(L+x)},
\end{equation}
where $A(q)$, $\Delta(k)$ represent the amplitude and phase shift,
respectively. With the form (\ref{scatt}) used in
(\ref{conti-app}), and after some careful study of the
$q$-dependence including that from $\psi_{q}^{\ast}(y)$, we notice
that the given amplitude takes the form
\begin{equation}
  \int_{-\infty}^{\infty}\!\!dq\ f(q,x,y)e^{ikL},
\end{equation}
where $f$ corresponds to some regular function in $q$. For $L$
very large, the order of magnitude for this integral can be
deduced with the help of the Riemann-Lesbegue lemma[6]: if the
$n$-th derivative of $f(q)$ satisfies the so-called Dirichlet
condition, then
\begin{equation}\label{RL}
  \int_{-\infty}^{\infty}\!\!f(q)e^{ikL}\sim
  O\left(\frac{1}{L^{n+1}}\right).
\end{equation}
Thus, for regular $f$, the integral should be smaller than any
power of $\frac{1}{L}$. This strongly suggests that the typical
large-$L$ behavior of the integral (\ref{RL}) is that of an
exponential suppression (i.e., vanishes like $e^{-\alpha L}$,
$\alpha$ being some positive constant). Remaining integrations
with respect to the variables $x$ and $y$ will not change this
order estimate in any significant way, and so $\kappa_{1}$ is
exponentially suppressed for large $L$. By analogous arguments one
may demonstrate that $\kappa_{2}$ and $\kappa_{2}'$, for large
$L$, are also exponentially small.

It is possible to give more precise large-$L$ dependences for the
factors $\kappa_{1}$, $\kappa_{2}$ and $\kappa_{2}'$ if a concrete
problem is considered. For instance, we can compute these factors
explicitlly when the problem is that of a pair of
$\delta$-function potentials as given in (\ref{delta-pot}). Then,
using the corresponding continuum wave-functions (see
(\ref{delta-eigen})), we obtain
\begin{eqnarray}
  \kappa_{1}&=&-\frac{2m\gamma_{1}\gamma_{2}}{\hbar^{2}}
  \int_{0}^{\infty}\!\!dq\frac{\langle x=L|q^{(1)}\rangle\langle q^{(1)}|y=0\rangle}
  {q^{2}+(\frac{m\gamma_{1}}{\hbar^{2}})^{2}}
  =\gamma_{2}\left(\frac{m\gamma_{1}L}{\hbar^{2}}-\frac{1}{2}\right)
  e^{-\frac{m\gamma_{1}L}{\hbar^{2}}},\label{factor1}\\
  \kappa_{2}&=&\frac{\gamma_{1}\gamma_{2}}{\gamma_{1}-\gamma_{2}}
  \left\{\frac{2\gamma_{2}}{\gamma_{1}+\gamma_{2}}
  e^{-\frac{m\gamma_{2}L}{\hbar^{2}}}-
  e^{-\frac{m\gamma_{1}L}{\hbar^{2}}}\right\},\label{factor2}\\
  \kappa_{2}'&=&\frac{4\hbar^{2}\gamma_{1}\gamma_{2}}{m}
  \left\{-\frac{\gamma_{2}}{(\gamma_{1}^{2}-\gamma_{2}^{2})^{2}
  }e^{-\frac{m\gamma_{2}L}{\hbar^{2}}}+\frac{1}{\gamma_{1}^{2}(\gamma_{1}-\gamma_{2})}
  \left(\frac{m\gamma_{1}L}{\hbar^{2}}+\frac{\gamma_{1}}{\gamma_{1}-\gamma_{2}}
  \right)e^{-\frac{m\gamma_{1}L}{\hbar^{2}}}\right\}.\label{factor3}
\end{eqnarray}
As anticipated, we see the exponential dependences on $L$ for
these factors. We also observe from the results
(\ref{factor1})-(\ref{factor3}) that, depending on the relative
magnitudes of $\gamma_{1}$ and $\gamma_{2}$, it may be just one
term that dominates the respective expression. Furthermore, as the
result (\ref{factor2}) for $\kappa_{2}$ is used in our formula
(\ref{2nd-eigen}), we notice that the second term (proportional to
$\delta E_{k}^{(1)}$) is negligible compared to the first. Hence,
for this example, we are allowed to write
\begin{equation}
  |\delta\phi_{k}\rangle^{(2)}=
  \hat{G}_{1}'\hat{V}_{1}\hat{G}_{2}\hat{V}_{2}|k\rangle=
  \kappa_{2}\sqrt{\frac{m\gamma_{1}}{\hbar^{2}}}
  e^{-\frac{m\gamma_{1}L}{\hbar^{2}}}\hat{G}_{1}'|x=0\rangle.
\end{equation}

\end{document}